**ARTICLE**

# On Enforcing Dyadic-type Homogeneous Binary Function Product Constraints in *MatBase*

*Christian Mancas*

*Mathematics and Computer Science Department, Ovidius University, Constanta, CT, 900 527, Romania*

**ABSTRACT**

Homogeneous binary function products are frequently encountered in the sub-universes modeled by databases, spanning from genealogical trees and sports to education and healthcare, etc. Their properties must be discovered and enforced by the software applications managing such data to guarantee plausibility. The (Elementary) Mathematical Data Model provides 17 types of dyadic-based homogeneous binary function product constraint categories. *MatBase*, an intelligent data and knowledge base management system prototype, allows database designers to simply declare them by only clicking corresponding checkboxes and automatically generates code for enforcing them. This paper describes the algorithms that *MatBase* uses for enforcing all 17 types of homogeneous binary function product constraint, which may also be employed by developers without access to *MatBase*.

## 1. Introduction

Very many database (db) sub-universes include homogeneous binary function products [1–4]. *Homogeneous binary function products* (hbfp) are, on one hand, cases of functions (i.e., of type $f \bullet g : D \to (C$ ∪ $NULLS)^2$, where *NULLS* is a distinguished countable set of *null values*), called *homogeneous* as the codomains of its two components are the same and, on the other, generalizations of dyadic relations (i.e., for which their canonical Cartesian projections might also take null values and their product must not be minimally one-to-one) [5].

As such, besides general function properties (i.e., minimal one-to-oneness/injectivity, ontoness/surjectivity, etc.), they may also enjoy all the 11 dyadic relation type ones (i.e., connectivity, reflexivity, irreflexivity, symmetry, asymmetry, transitivity, intransitivity, Euclideanity, inEuclideanity, equivalence, and acyclicity).

For example, *Husband • Wife* : *MARRIAGES* → *PERSONS*$^2$, *Mother • Spouse* : *PERSONS* → (*PERSONS* ∪ NULLS)$^2$, *Father • Spouse* : *PERSONS* → (*PERSONS* ∪ NULLS)$^2$, and *Mother • Father* : *PERSONS* → (*PERSONS* ∪ NULLS)$^2$ are hbfps frequently encountered in db schemes. All 4 are:

✓ irreflexive (as nobody may be his own spouse, no parent may be somebody's spouse, and no mother may also be a father, respectively);

✓ asymmetric (as no husband may be a wife or vice-versa, and whenever $y$ is $x$'s parent and $z$ is his/her spouse, neither $z$ may be $x$'s parent, nor $y$ may be $x$'s spouse, and whenever $y$ is $x$'s mother and $z$ is his/her father, neither $z$ may be $x$'s mother, nor $y$ may be $x$'s father, respectively);

✓ acyclic (as no husband may be a wife or vice-versa, no parent may be somebody's spouse, and no mother may also be a father, respectively); and

✓ inEuclidean (same as for acyclicity).

As both $f$ and $g$ may take null values (e.g., there are single persons, as well as persons without known fathers or/and mothers), the (Elementary) Mathematical Data Model ((E)MDM) [2,4] considers 6 additional types of dyadic-type hbfp constraints [6–8], namely (where *iff* is the abbreviation for "if and only if"):

o *null-reflexivity*, iff, for any element $y$ of $C$, there is at least one element $x$ of $D$ such that either $f(x) = y = g(x)$, or $f(x) = y$ and $g(x) \in$ NULLS, or $g(x) = y$ and $f(x) \in$ NULLS;

o *null-identity*, iff, for any element $x$ of $D$, either $f(x) = g(x)$, or $f(x) = x$ and $g(x) \in$ NULLS, or $g(x) = x$ and $f(x) \in$ NULLS, or $f(x) \in$ NULLS and $g(x) \in$ NULLS;

o *null-symmetry*, iff, for any elements $u$ and $v$ of $C$ for which there is an $x$ of $D$ such that $f(x) = u$ and $g(x) = v$, there is at least one element $y$ of $D$ such that either $f(y) = v$ and $g(y) = v$, or $f(y) = u$ and $g(y) \in$ NULLS, or $g(y) = v$ and $f(y) \in$ NULLS;

o *null-transitivity*, iff, for any elements $u$, $v$ and $w$ of $C$ for which there are elements $x$ and $y$ of $D$ such that $f(x) = u$, $g(x) = v$, $f(y) = v$, and $g(y) = w$, there is at least one element $z$ of $D$ such that either $f(z) = u$ and $g(z) = w$, or $f(z) = u$ and $g(z) \in$ NULLS, or $g(z) = w$ and $f(z) \in$ NULLS;

o *null-Euclideanity*, iff, for any elements $u$, $v$ and $w$ of $C$ for which there are elements $x$ and $y$ of $D$ such that $f(x) = u$, $g(x) = v$, $f(y) = u$, and $g(y) = w$, there is at least one element $z$ of $D$ such that either $f(z) = v$ and $g(z) = w$, or $f(z) = v$ and $g(z) \in$ NULLS, or $g(z) = w$ and $f(z) \in$ NULLS;

o *null-equivalence*, iff $f • g$ is both null-reflexive and null-Euclidean.

On one hand, as with any other constraint (business rule), failing to enforce any of the above ones could lead to storing unplausible data in the corresponding db (e.g., for some persons $x$ and $y$, *Mother*($x$) = $y$ = *Spouse*($x$), or/and *Mother*($x$) = $y$ ∧ *Spouse*($y$) = $x$, etc.).

On the other hand, as connectivity, reflexivity, symmetry, transitivity, and Euclideanity constraints are of type tuple generating, enforcing them is also saving time for end-users, as the corresponding hbfp values might be automatically generated by the db software applications managing that data.

Of course, the dyadic-type hbfp constraint types are not enough to guarantee data plausibility, not even for *PERSONS*: as usual, all other existing constraints in the corresponding sub-universe should also be enforced. For example, *Mother acyclic* (No woman may be the mother of one of her ascendants or grandchildren or descendants of at least 2 generations.), *Spouse null-symmetric* (If $y$ is $x$'s spouse, then $x$ is $y$'s spouse.), etc.

Unfortunately, while, for example, uniqueness may be enforced by almost any commercial Database Management System (DBMS) with unique indexes, no such DBMS may enforce dyadic-type hbfp constraints. Consequently, developers must enforce them within the software applications that manage corresponding dbs (through either extended SQL triggers or event-driven methods of high-level programming languages embedding SQL).

*MatBase* [2–4,6–10] is an intelligent prototype data and knowledge base management system, based on both the (E)MDM, the Entity-Relationship (E-R) Data Model (E-RDM) [1,11,12], the Relational Data Model (RDM) [1,13,14], and Datalog¬ [4,15,16].

Fortunately, *MatBase* provides, through its (E)MDM interface, a very user-friendly experience for db architects (e.g., for the *Husband • Wife*, *Mother • Father*, *Mother • Spouse*, and *Father • Spouse* of *PERSONS* above, you only need to click their corresponding *inEuclidean* and *Acyclic* checkboxes, as their other 2 properties are implied by it, so redundant) and its associated code-generating power, which is both constructing underlying db tables, standard MS Windows forms for them, as well as event-driven code in their classes for enforcing the corresponding constraints that cannot be enforced by the available DBMSes.

As such, MatBase not only saves developing time but also saves testing and debugging time, which promotes the 5th programming language generation—modelling as programming [9,10,16]. This paper presents the pseudocode algorithms used by *MatBase* to automatically generate code for enforcing dyadic-type hbfp constraints.

Other approaches related to the (E)MDM are based on business rules management (BRM) [17–22] and their corresponding implemented systems (BRMS) and process managers (BPM), like the IBM Operational Decision Manager [23], IBM Business Process Manager [24], Red Hat Decision Manager [25], Agiloft Custom Workflow/BPM [26], etc.

They are generally based on XML (but also on the Z notation, the Business Process Execution Language, the Business Process Modeling Notation, the Decision Model and Notation models, Drools Rule Language files, guided decision tables, or the Semantics of Business Vocabulary and Business Rules).

This is the only other field of endeavor trying to systematically deal with business rules, even informally. However, this is not done at the db design level, but at the software application one, and without providing automatic code generation.

From this perspective, (E)MDM also belongs to the panoply of tools expressing business rules and *MatBase* is also a BRMS, but a formal, automatic code-generating one.

## 2. Materials and methods

Let $f \bullet g : D \rightarrow (C \cup \text{NULLS})^2$ be an arbitrary hbfp. For enforcing hbfp dyadic type constraints on $f \bullet g$, $C$ and $D$ must have Graphic User Interface (GUI) forms (associated to their corresponding tables) and (see **Figure 1**):

➢ Classes $C$ and $D$ of these forms must contain private *AfterInsert*($x$) and *AfterInsert*($f$, $g$) methods, respectively;

➢ Moreover, class $D$ must contain:

o The definition of two private numeric variables *fOldValue* and *gOldValue*;

o A private method *Current*($x$, $f$, $g$), to be called each time the cursor of the $D$'s form enters a new element (line, row, record) $x$ of its underlying data;

o A private method *BeforeInsert*($x$, $f$, $g$), to be called each time end-users ask for adding a new element $x$ to $D$;

o A private method *BeforeUpdate*($x$, $f$, $g$), to be called each time a new or existing element $x$ of its underlying data whose values for columns <$f$, $g$> were <*fOldValue*, *gOldValue*> and that were then modified to <$u$, $v$> is about to be saved in the db;

o A private method *AfterUpdate*($x$, $f$, $g$), to be called each time an existing element $x$ of its underlying data whose values for columns <$f$, $g$> were <*fOldValue*, *gOldValue*> were then modified to <$u$, $v$> and successfully saved to the db;

o A private method *Delete*($x$, $f$, $g$), to be called each time end-users ask for the deletion of an existing $x$ element of its underlying data;

o A private method *AfterDelSuccess*($x$, $f$, $g$), to be called each time an existing element $x$ of its underlying data whose values for columns <$f$, $g$> were <*fOldValue*, *gOldValue*> was successfully deleted from the db.

All these methods and variables are automatically generated by *MatBase* the first time it needs them.

Code insertion in all these methods is always done immediately before their blank code line (see **Figure 1**).

Moreover, *MatBase* has in its public *General* library the function definition *IsCycle*, based on the pseudocode algorithm from **Figure 2**, needed for enforcing hbfp acyclicities (which is a variant of the *computeDyadicRelInstantiationTransClosure* one, based on the pseudocode algorithm shown in Figure 21 from Mancas [27]).

| // class C | // class D |
|---|---|
| **Method** *AfterInsert(x)* | **Method** *AfterInsert(x, f, g)* |
| // class D<br>int fOldValue, gOldValue,<br>Boole INS; | **Method** *BeforeUpdate(x, f, g)*<br>Boole Cancel = False; |
| **Method** *Current(x, f, g)*<br>fOldValue = f;<br>gOldValue = g;<br>INS = False; | *if Cancel then* deny saving <f, g> to D;<br>**Method** *AfterUpdate(x, f, g)*<br>*if INS then* requery D; |
| | **Method** *Delete (x, f, g)*<br>Boole Cancel = False; |
| **Method** *BeforeInsert(x, f, g)*<br>Boole Cancel = False; | *if Cancel then* deny deletion of <f, g> from D; |
| *if Cancel then* deny inserting <f, g> into D; | **Method** *AfterDelSuccess(x, f, g)* |

**Figure 1.** Initially generated pseudocode for classes *C* and *D*.

## 2.1 Enforcing connectivity constraints

Generally, a hbfp $f \bullet g : D \rightarrow C^2$ is *connected* whenever, for any values $y, z \in C$, $y \neq z$, there is a value $x$ in $D$ such that $y == f(x)$ and $z == g(x)$ or $z == f(x)$ and $y == g(x)$. According to this definition, enforcing such constraints for $f \bullet g$ requires that:

(i). Each time a new element $x$ is added to $C$, pairs $<x, y>$ or $<y, x>$ must be automatically added to $f \bullet g$'s graph, for any other element $y$ of $C$. Moreover, whenever $f \bullet g$ is also symmetric, both these pairs should be added.

(ii). Each time a pair $<x, y>$ of $f \bullet g$'s graph, $y \neq x$, is modified in $<u, v>$, with either $u \neq x$ or $v \neq y$ and $u, v \notin$ NULLS, and there is no $<y, x>$ or $<y, >$ in $f \bullet g$'s graph, then either $<x, y>$ or $<y, x>$ must be automatically added to $f \bullet g$'s graph. Moreover, whenever $f \bullet g$ is also symmetric, such pair may be modified to only $<x, >$ or $<, y>$.

(iii). No pair $<x, y>$ of $f \bullet g$'s graph, $y \neq x$, $x$, $y$ not nulls, should be deleted, if there is no pair $<y, x>$ or $<y, >$ or $<, x>$ in $f \bullet g$'s graph. Moreover, whenever $f \bullet g$ is also symmetric, no such pair should ever be deleted, unless both $x$ and $y$ are nulls.

Consequently, *MatBase* adds the pseudocode algorithms from **Figure 3** to the ones in **Figure 1**.

## 2.2 Enforcing reflexivity and null-reflexivity constraints

Generally, a hbfp $f \bullet g : D \rightarrow C^2$ is *reflexive* whenever, for any value $y \in C$, there is a value $x$ in $D$ such that $y = f(x) = g(x)$. In other words, it is reflexive whenever its instance (i.e., $Im(f \bullet g)$) contains the first diagonal of the product between the union of the instances of its members with itself (i.e., $(Im(f) \cup Im(g))^2$, where $Im(f) = \{y \mid \exists x \in D, y = f(x)\} \subseteq C$).

According to this definition, enforcing such constraints for $f \bullet g$ requires that:

(i). Each time a new element $x$ is added to $C$, the system must automatically add the reflexive pair $<x, x>$ to $Im(f \bullet g)$.

(ii). No pair $<x, x>$ may be modified to anything else than $<x, >$ or $<, x>$; dually, the only accepted modification of a pair $<x, >$ or $<, x>$ is $<x, x>$.

```
public Boolean function IsCycle(text D, text #D, int x, text f, int u, text g, int v)
// returns True if <u=f(x), v=g(x)> closes at least one cycle in f • g's graph and False otherwise;
// based on the algorithm for computing instantiations of hbfp transitivity closures;
Input: functions (columns) f and g defined on (table) D and their values u and v for x, the one
       of the autonumber #D (object id, primary key) for D's current element (row);
Strategy:
IsCycle = False;                                  // no cycle found yet
if u ∈ NULLS or v ∈ NULLS then IsCycle = True;    // null values may not close cycles
elseif u == v then IsCycle = True;                // reflexivity cycle found!
elseif ∃y∈D such that f(y) == v and g(y) == u then IsCycle = True;  // symmetry cycle found!
else                                              // check for higher arity cycles
   if existsTable("TransClosure") then execute("DROP TABLE TransClosure");
   execute("CREATE TABLE TransClosure(x COUNTER, [Level] INT, d INT, a INT)");
   int oldCard = 0;                               // TransClosure is empty
   // initialize TransClosure with the not null elements of the f • g's graph, except for x
   execute("INSERT INTO TransClosure SELECT 1 AS [Level], " & f & ", " & g &
           " FROM [" & D & "] WHERE " & #D & " <>" & x & " AND NOT [" & f & "] IS " 
           & "NULL AND NOT [" & g & "] IS NULL");
   execute("DELETE FROM TransClosure WHERE x IN (SELECT x FROM (SELECT x, " 
           & "d, a FROM TransClosure AS Tmp GROUP BY d, a HAVING Count(*) > 1 AND"
           & " a = TransClosure.a)) AS TCdups WHERE x NOT IN (SELECT MIN(x) FROM "
           & "TCdups GROUP BY d, a)");           // delete duplicates
   int Card = execute("SELECT COUNT(*) FROM TransClosure");
   int level = 2;
   while Card ≠ oldCard and not IsCycle
      oldCard = Card;                            // avoid infinite looping
      // add to the transitive closure the next level elements
      execute("INSERT INTO TransClosure SELECT " & level & ", TransClosure.d, [" & D &
              "].[" & g & "] FROM [" & D & "] INNER JOIN TransClosure ON TransClosure.a = ["
              & D & "].[" & f & "] WHERE NOT [" & D & "].[" & g & "] IS NULL AND "
              & "[Level] =" & level – 1 & " AND #D & " <>" & x);
      Card = execute("SELECT Count(*) FROM TransClosure WHERE d = " & v &
              " AND a =" & u)   // how many <v, u> pairs were computed in this step?
      if Card > 0 then IsCycle = True;           // cycle of length level found!
      else       // delete duplicates, set next level, and compute new TransClosure's cardinal
         execute("DELETE FROM TransClosure WHERE x IN (SELECT x FROM (SELECT"
              & " x, d, a FROM TransClosure AS Tmp GROUP BY d, a HAVING Count(*) "
              & "> 1 AND a = TransClosure.a)) AS TCdups WHERE x NOT IN "
              & "(SELECT MIN(x) FROM TCdups GROUP BY d, a)");
         level = level + 1;
         Card = execute("SELECT COUNT(*) FROM TransClosure");
      end if;
   end while;
end if;
end function IsCycle;
```

**Figure 2.** The pseudocode of the function *IsCycle* from the *MatBase*'s *General* library.

```
// code added to method AfterInsert of class C
// f • g connected
loop for all y ∈ C, y ≠ x
    add <x, y> to Im(f • g); INS = True; if f • g symmetric then add <y, x> to Im(f • g);
end loop;
// code added to method BeforeUpdate of class D
// f • g connected
if not Cancel ∧ ¬ (f == fOldValue ∧ g ∈ NULLS ∨ f ∈ NULLS ∧ g == gOldValue) ∧ f • g
    symmetric then Cancel = True; display "Request rejected: f • g connected and symmetric!";
end if;
// code added to method AfterUpdate of class D
// f • g connected
if f • g ¬symmetric ∧ fOldValue ≠ gOldValue ∧ (f ≠ fOldValue ∨ g ≠ gOldValue) ∧
    f,g ∉ NULLS ∧ <gOldValue, fOldValue> ∉ Im(f • g) then
        add <gOldValue, fOldValue> to Im(f • g); INS = True;
end if;
// code added to method Delete of class D
// f • g connected
if not Cancel ∧ f • g symmetric then if f ∉ NULLS ∨ g ∉ NULLS then Cancel = True;
else if not Cancel ∧ f ∉ NULLS ∧ g ∉ NULLS ∧ f ≠ g ∧ (¬∃x∈D)(<g(x), f(x)> ∈ Im(f • g) ∨
    <g(x), > ∈ Im(f • g) ∨ <, f(x)> ∈ Im(f • g)) then Cancel = True;
end if;
if Cancel then t = "Request rejected: f • g connected";
    if f • g symmetric then t = t & " and symmetric";
    t = t & "!"; display t;
end if;
```

**Figure 3.** Code added to the methods from Figure 1 when $f \bullet g$ is connected.

(iii). Deletion of $\langle x, x \rangle$ may be accepted only if $Im(f \bullet g)$ also contains $\langle x, \rangle$ or $\langle , x \rangle$; deletion of $\langle x, \rangle$ may be accepted only if $Im(f \bullet g)$ also contains $\langle x, x \rangle$ or $\langle , x \rangle$; deletion of $\langle , x \rangle$ may be accepted only if $Im(f \bullet g)$ also contains $\langle x, x \rangle$ or $\langle x, \rangle$.

Consequently, *MatBase* adds the pseudocode algorithms from **Figure 4** to the corresponding methods from **Figure 1**, but only when $f \bullet g$ is not null-identical as well (when, according to the algorithms for the coherence and minimality of the constraint sets [2,4,7,8], reflexivity is redundant).

### 2.3 Enforcing null-identity constraints

Note that for hbfps the case of the *identity* (i.e., $(\forall x \in D)(f(x) = g(x))$) is not semantically interesting at all, as it means that you would request, in the end, that two columns of a same table to always have identical values, i.e., maintain a redundant copy of a column.

However, null-identity might be of use, e.g., when it is desired to store the same values for two distinct properties of an object set whenever it is applicable for both, but some of them might not be applicable to both. Obviously, any null-identical hbfp is also null-reflexive.

According to the null-identity definition, enforcing such constraints for $f \bullet g$ requires that only pairs of type $\langle y, y \rangle, \langle , y \rangle, \langle y, \rangle$, or $\langle , \rangle, y \notin$ NULLS, may be saved to $Im(f \bullet g)$'s graph.

Consequently, *MatBase* adds the pseudocode algorithm from **Figure 5** to the corresponding method from **Figure 1**.

```
// code added to method AfterInsert of class C
// f • g reflexive
add <x, x> to Im(f • g); INS = True;
// code added to method BeforeUpdate of class D
// f • g reflexive
if not Cancel ∧ fOldValue == gOldValue ∧ fOldValue ∉ NULLS ∧ gOldValue ∉ NULLS ∧
    (f ∉ NULLS ∧ f ≠ fOldValue ∨ g ∈ NULLS) ∧
    (f ∉ NULLS ∨ g ≠ gOldValue ∧ g ∉ NULLS) then Cancel = True;
elseif not Cancel ∧ (fOldValue ∉ NULLS ∧ gOldValue ∈ NULLS ∨ gOldValue ∉ NULLS ∧
    fOldValue ∈ NULLS) ∧ (f ≠ fOldValue ∧ f ∈ NULLS ∨ g ≠ fOldValue ∨ g ∈ NULLS)
    then Cancel = True;
if Cancel then display "Request rejected: f • g reflexive!";
// code added to method Delete of class D
// f • g reflexive
if not Cancel ∧ f == g ∧ f ∉ NULLS ∧ then
    if (¬∃y∈D)(<f(y) == f, > ∈ Im(f • g) ∨ <, g(y) == g> ∈ Im(f • g)) then Cancel = True;
elseif not Cancel ∧ f ∉ NULLS ∧ g ∈ NULLS then
    if (¬∃y∈D)(<f(y) == f, g(y) == f> ∈ Im(f • g) ∨ <, g(y) == f> ∈ Im(f • g)) then Cancel = True;
elseif not Cancel ∧ f ∈ NULLS ∧ g ∉ NULLS then
    if (¬∃y∈D)(<f(y) == g, g(y) == g> ∈ Im(f • g) ∨ <f(y) == g,> ∈ Im(f • g)) then Cancel = True;
if Cancel then display "Request rejected: f • g reflexive!";
```

**Figure 4.** Code added to the methods from Figure 1 when $f \bullet g$ is reflexive.

```
// code added to method BeforeUpdate of class D
// f • g null-identical
if not Cancel ∧ ¬(f ∈ NULLS ∧ g ∈ NULLS ∨ f == g ∨ f ∈ NULLS ∧ g ∉ NULLS ∨ f ∉
    NULLS ∧ g ∈ NULLS) then Cancel = True; display "Request rejected: f • g null-identical!";
end if;
```

**Figure 5.** Code added to the methods from Figure 1 when $f \bullet g$ is null-identical.

### 2.4 Enforcing irreflexivity constraints

Generally, a hbfp $f \bullet g : D \rightarrow C^2$ is *irreflexive* whenever there is no value $x$ in $D$ such that $y = f(x) = g(x)$. According to this definition, enforcing such constraints for $f \bullet g$ requires that each time a pair $\langle x, x \rangle$ (be it new or obtained by modifying an existing $\langle u, v \rangle$) is about to be saved in the db $f \bullet g$'s image, saving must be canceled.

Consequently, *MatBase* adds the pseudocode algorithm from **Figure 6** to the corresponding method from **Figure 1**, but only when $f \bullet g$ is neither asymmetric, nor intransitive, nor inEuclidean as well (cases in which, according to the algorithms for the coherence and minimality of the constraint sets [2,4,7,8], irreflexivity is redundant).

```
// code added to method BeforeUpdate of class D
// f • g irreflexive
if not Cancel ∧ f == g then Cancel = True; display "Request rejected: f • g irreflexive!"; end if;
```

**Figure 6.** Code added to the methods from Figure 1 when $f \bullet g$ is irreflexive.

## 2.5 Enforcing symmetry and null-symmetry constraints

Generally, a hbfp $f \bullet g : D \to C^2$ is *symmetric* whenever, for any pair $<u, v> \in Im(f \bullet g)$, there is a pair $<v, u> \in Im(f \bullet g)$. According to this definition, enforcing such constraints for $f \bullet g$ not connected (the case $f \bullet g$ connected is dealt with in subsection 2.1) requires that:

(i). Each time a pair $<x, y>$, x ≠ y, is added to $Im(f \bullet g)$, a pair $<y, x>$ must automatically be added to $Im(f \bullet g)$ as well.

(ii). Each time a pair $<x, x>$ of $Im(f \bullet g)$ is modified in $<u, v>$, with $u \neq v$ and either $u \neq x$ or $v \neq x$, then $<v, u>$ must automatically be added to $Im(f \bullet g)$; each time a pair $<x, y>$ of $Im(f \bullet g)$, $y \neq x$, is modified in $<u, v>$, with $u \neq v$ and either $u \neq x$ or $v \neq y$, then $<y, x>$ must automatically be replaced in $Im(f \bullet g)$ by $<v, u>$, whenever $f \bullet g$ is not connected; and each time a pair $<x, y>$ of $Im(f \bullet g)$, $y \neq x$, is modified in $<u, u>$ and either $u \neq x$, or $u \neq y$, then $<y, x>$ must automatically be deleted from $Im(f \bullet g)$, whenever $f \bullet g$ is not connected.

(iii). Each time a pair $<x, y>$ of $Im(f \bullet g)$, $y \neq x$, is deleted, then $<y, x>$ must automatically be deleted from $Im(f \bullet g)$ as well, whenever $f \bullet g$ is not connected.

Consequently, *MatBase* adds the pseudocode algorithms from **Figure 7** to the corresponding methods from **Figure 1**.

In Mancas [4,28] it is shown that $Im(f \bullet g) = D^2$, so computable, whenever $f \bullet g$ is both symmetric, reflexive, and connected.

```
// code added to method AfterInsert of class D
// f • g symmetric
if f • g not connected ∧ f ≠ g then add <g,f> to Im(f • g); INS = True; end if;
// code added to method AfterUpdate of class D
// f • g symmetric
if f • g not connected ∧ (f ≠ fOldValue ∨ g ≠ gOldValue) then
    if f ≠ g ∧ fOldValue ≠ gOldValue then replace <gOldValue, fOldValue> by <g, f>;
    elseif f == g ∧ fOldValue ≠ gOldValue then delete <gOldValue, fOldValue> from Im(f • g);
    elseif f ≠ g ∧ fOldValue == gOldValue then add <g, f> to Im(f • g); INS = True;
    end if;
end if;
// code added to method AfterDelSuccess of class D
// f • g symmetric
if f • g not connected ∧ f ≠ g ∧ fOldValue ≠ gOldValue then
    replace <gOldValue, fOldValue> by <g, f>;
```

**Figure 7.** Code added to the methods from Figure 1 when $f \bullet g$ is symmetric.

## 2.6 Enforcing asymmetry constraints

Generally, a hbfp $f \bullet g : D \to C^2$ is *asymmetric* whenever, for any pair $<u, v> \in Im(f \bullet g)$, there is no pair $<v, u> \in Im(f \bullet g)$. According to this definition, enforcing such constraints for $f \bullet g$ requires that:

(i) Each time a pair $<x, y>$, $x \neq y$, is about to be added to $Im(f \bullet g)$, this must be rejected whenever a pair $<y, x>$ exists in $Im(f \bullet g)$.

(ii) Each time a pair $<x, y>$ of $Im(f \bullet g)$ is modified in $<u, v>$, with $u \neq v$ and either $u \neq x$ or $v \neq y$, this must be rejected whenever a pair $<v, u>$ exists in $Im(f \bullet g)$.

Consequently, *MatBase* adds the pseudocode algorithms from **Figure 8** to the corresponding methods from **Figure 1**, but only when $f \bullet g$ is neither acyclic, nor irreflexive and transitive as well (cases in which, according to the algorithms for the coherence and minimality of the constraint sets [2,4,7,8], asymmetry is redundant), nor connected (as asymmetry implies not connectivity—see Mancas [4,28]).

```
// code added to method BeforeInsert of class D
// f • g asymmetric
if not Cancel ∧ f ≠ g ∧ f ∉ NULLS ∧ g ∉ NULLS then
    if <g, f> ∈ Im(f • g) then Cancel = True; display "Request rejected: f • g asymmetric!"; end if;
end if;
// code added to method BeforeUpdate of class D
// f • g asymmetric
if not Cancel ∧ f ≠ g ∧ f ∉ NULLS ∧ g ∉ NULLS ∧ (f ≠ fOldValue ∨ g ≠ gOldValue) then
    if <g, f> ∈ Im(f • g) then Cancel = True; display "Request rejected: f • g asymmetric!"; end if;
end if;
```

**Figure 8.** Code added to the methods from Figure 1 when $f \bullet g$ is asymmetric.

## 2.7 Enforcing transitivity and null-transitivity constraints

Generally, a hbfp $f \bullet g : D \to C^2$ is *transitive* whenever, for any pairs $<u, v>, <v, w> \in Im(f \bullet g)$, there is a pair $<u, w> \in Im(f \bullet g)$. According to this definition, enforcing such constraints for $f \bullet g$ requires that:

(i). Each time a pair $<x, y>$, $x \neq y$, is added to $Im(f \bullet g)$ and $Im(f \bullet g)$ contains a pair $<y, z>$, $z \neq y$, a pair $<x, z>$ must automatically be added to $Im(f \bullet g)$ as well, iff it does not exist already.

(ii). Each time a pair $<x, z>$ of $Im(f \bullet g)$ is mod-ified in <u, v>, with either u ≠ x or v ≠ z, and there is at least a y in C such that both <x, y> and <y, z> belong to Im(f • g), then modification of <x, z> must be rejected; each time a pair <x, x> of Im(f • g) is modified in <u, v>, with u ≠ v and either u ≠ x or v ≠ x, and there is at least a y in C such that either <u, y> or <y, v> are in Im(f • g), then either <y, v> or <u, y> must automatically be added to Im(f • g), iff they do not exist already.

(iii). Each time a pair <x, z> of Im(f • g) is about to be deleted and there is at least a y in C such that both <x, y> and <y, z> belong to Im(f • g), then deletion of <x, z> must be rejected.

Consequently, *MatBase* adds the pseudocode algorithms from **Figure 9** to the corresponding methods from **Figure 1**, but only when f • g is neither connected and symmetric, nor Euclidean as well (cases in which, according to the algorithms for the coherence and minimality of the constraint sets [2,4,7,8], transitivity is redundant), nor intransitive as well (case in which it adds the algorithms from **Figure 11**).

## 2.8 Enforcing intransitivity constraints

Generally, a hbfp f • g : D → $C^2$ is *intransitive* whenever, for any pairs <u, v>, <v, w> ∈ Im(f • g), there is no pair <u, w> ∈ Im(f • g). According to this definition, enforcing such constraints for f • g requires that:

(i). Each time a pair <x, z> is about to be added to Im(f • g) and there are at least two pairs <x, y> and <y, z> stored by Im(f • g), then adding <x, z> to Im(f • g) must be rejected.

(ii). Each time a pair <u, v> of Im(f • g) is modified in <x, z>, with either u ≠ x or v ≠ z, and there is at least a y in C such that both <x, y> and <y, z> belong to Im(f • g), with y ≠ x and y ≠ z, then modification of <u, v> must be rejected.

Consequently, *MatBase* adds the pseudocode algorithms from **Figure 10** to the corresponding methods from **Figure 1**, but only when f • g is not symmetric and inEuclidean as well (case in which, according to the algorithms for the coherence and minimality of the constraint sets [2,4,7,8], intransitivity is redundant) and when it is not also transitive.

```
// code added to method AfterInsert of class D
// f • g transitive and not intransitive
if f ≠ g ∧ f ∉ NULLS ∧ g ∉ NULLS then
    loop for all <g, z> ∈ Im(f • g), g ≠ z ∧ z ∉ NULLS
        if <f, z> ∉ Im(f • g) then add <f, z> to Im(f • g); INS = True; end if;
    end loop;
end if;
// code added to method BeforeUpdate of class D
// f • g transitive and not intransitive
if not Cancel ∧ (f ≠ fOldValue ∨ g ≠ gOldValue) then
    if fOldValue ≠ gOldValue then
        if ∃z∈C such that <fOldValue, z> ∈ Im(f • g) ∧ <z, gOldValue> ∈ Im(f • g) then
            Cancel = True; display "Request rejected: f • g transitive!";
        end if;
    elseif f ≠ g then
        loop for all <f, z> ∈ Im(f • g), f ≠ z ∧ z ∉ NULLS
            if <z, g> ∉ Im(f • g) then add <z, g> to Im(f • g); INS = True; end if;
        end loop;
        loop for all <z, g> ∈ Im(f • g), g ≠ z ∧ z ∉ NULLS
            if <f, z> ∉ Im(f • g) then add <f, z> to Im(f • g); INS = True; end if;
        end loop;
    end if;
end if;
// code added to method Delete of class D
// f • g transitive and not intransitive
if not Cancel ∧ ∃z∈C such that <f, z> ∈ Im(f • g) ∧ <z, g> ∈ Im(f • g) then
    Cancel = True; display "Request rejected: f • g transitive!";
end if;
```

**Figure 9.** Code added to the methods from Figure 1 when f • g is transitive and not intransitive.

```
// code added to method BeforeInsert of class D
// f • g intransitive and not transitive
if not Cancel ∧ ∃z∈C such that <f, z> ∈ Im(f • g) ∧ <z, g> ∈ Im(f • g) then
    Cancel = True; display "Request rejected: f • g intransitive!";
end if;
// code added to method BeforeUpdate of class D
// f • g intransitive and not transitive
if not Cancel ∧ ∃z∈C such that <f, z> ∈ Im(f • g) ∧ <z, g> ∈ Im(f • g) ∧ f ∉ NULLS ∧
    g ∉ NULLS then Cancel = True; display "Request rejected: f • g intransitive!";
end if;
```

**Figure 10.** Code added to the methods from Figure 1 when f • g is intransitive and not transitive.

Indeed, unlike reflexivity and irreflexivity or symmetry and asymmetry, which are duals of each other, respectively, transitivity and intransitivity are orthogonal to each other, i.e., there may be relations that are both transitive and intransitive iff, for any pair <u, v> ∈ Im(f • g), u ≠ v, there is no pair <v, w> ∈ Im(f • g), w ≠ v.

According to this definition, enforcing this pair of constraints for f • g requires that:

(i). Each time a pair <v, w> is about to be added to Im(f • g) and there is a pair <u, v> stored by Im(f • g), with u ≠ v and w ≠ v, then adding <x, w> to Im(f • g) must be rejected.

(ii). Each time a pair <x, y> of Im(f • g) is modified in <v, w>, with v ≠ w, and there is at least a u in C such that <u, v> belongs to Im(f • g), with u ≠ v, then modification of <x, y> must be rejected.

Consequently, *MatBase* adds the pseudocode algorithms from **Figure 11** to the corresponding meth-ods from **Figure 1** whenever $f \bullet g$ is both transitive and intransitive.

```
// code added to method BeforeInsert of class D
// f • g transitive and intransitive
if not Cancel ∧ f ≠ g ∧ ∃u∈C such that <u, f> ∈ Im(f • g), u ≠ f then
    Cancel = True; display "Request rejected: f • g both transitive and intransitive!";
end if;
// code added to method BeforeUpdate of class D
// f • g transitive and intransitive
if not Cancel ∧ f ≠ g ∧ ∃u∈C such that <u, f> ∈ Im(f • g) , u ≠ f then
    Cancel = True; display "Request rejected: f • g both transitive and intransitive!";
end if;
```

**Figure 11.** Code added to the methods from Figure 1 when $f \bullet g$ is both transitive and intransitive.

## 2.9 Enforcing Euclideanity and null-Euclideanity constraints

Generally, a hbfp $f \bullet g : D \rightarrow C^2$ is *Euclidean* whenever, for any pairs $<u, v>$, $<u, w> \in Im(f \bullet g)$, $v \neq w$, there are pairs $<v, w>$, $<w, v> \in Im(f \bullet g)$. According to this definition, enforcing such constraints for $f \bullet g$ requires that:

(i). Each time a pair $<x, y>$ is added to $Im(f \bullet g)$ and $Im(f \bullet g)$ contains a pair $<x, z>$, pairs $<y, z>$ and $<z, y>$ must automatically be added to $Im(f \bullet g)$ as well, iff they do not exist already.

(ii). Each time a pair $<y, z>$ or $<z, y>$ of $Im(f \bullet g)$ is modified in $<u, v>$, with either $u \neq y$ or $u \neq z$ or $v \neq z$ or $v \neq y$, respectively, and there is at least a $x$ in $C$ such that both $<x, y>$ and $<x, z>$ belong to $Im(f \bullet g)$, then modification of $<y, z>$ or $<z, y>$ must be rejected.

(iii). Each time a pair $<y, z>$ or $<z, y>$ of $Im(f \bullet g)$ is about to be deleted and there is at least a $x$ in $C$ such that both $<x, y>$ and $<x, z>$ belong to $Im(f \bullet g)$, then deletion of $<y, z>$ or $<z, y>$ must be rejected.

Consequently, *MatBase* adds the pseudocode algorithms from **Figure 12** to the corresponding methods from **Figure 1**, but only when $f \bullet g$ is neither connected and symmetric, nor transitive and symmetric as well (case in which, according to the algorithms for the coherence and minimality of the constraint sets [2,4,7,8], Euclideanity is redundant) and when it is not also inEuclidean.

In Mancas [4,28] it is also shown that any Euclidean $f \bullet g$ is also symmetric and transitive and may not be either acyclic or asymmetric or irreflexive or in-transitive; moreover, any Euclidean $f \bullet g$ that is also reflexive and connected is computable, as the corresponding $Im(f \bullet g)$ is equal to $D^2$.

```
// code added to method AfterInsert of class D
// f • g Euclidean and not inEuclidean
if f ≠ g ∧ f ∉ NULLS ∧ g ∉ NULLS then
    loop for all <f, z> ∈ Im(f • g), f ≠ z ∧ z ∉ NULLS
        if <g, z> ∉ Im(f • g) then add <g, z> to Im(f • g); INS = True; end if;
        if <z, g> ∉ Im(f • g) then add <z, g> to Im(f • g); INS = True; end if;
    end loop;
end if;
// code added to method BeforeUpdate of class D
// f • g Euclidean and not inEuclidean
if not Cancel ∧ (f ≠ fOldValue ∨ g ≠ gOldValue) then
    if ∃z∈C such that <z, fOldValue> ∈ Im(f • g) ∧ <z, gOldValue> ∈ Im(f • g) then
        Cancel = True; display "Request rejected: f • g Euclidean!";
    end if;
end if;
// code added to method Delete of class D
// f • g Euclidean and not inEuclidean
if not Cancel ∧ ∃z∈C such that <z, f> ∈ Im(f • g) ∧ <z, g> ∈ Im(f • g) then
        Cancel = True; display "Request rejected: f • g Euclidean!";
end if;
```

**Figure 12.** Code added to the methods from Figure 1 when $f \bullet g$ is Euclidean and not inEuclidean.

## 2.10 Enforcing inEuclideanity constraints

Generally, a hbfp $f \bullet g : D \rightarrow C^2$ is *inEuclidean* whenever, for any pairs $<u, v>$, $<u, w> \in Im(f \bullet g)$, $v \neq w$, there are no pairs $<v, w>$ or $<w, v> \in Im(f \bullet g)$. According to this definition, enforcing such constraints for $f \bullet g$ requires that:

(i). Each time a pair $<y, z>$ or $<z, y>$ is about to be added to $Im(f \bullet g)$ and there are at least two pairs $<x, y>$ and $<x, z>$ stored by $Im(f \bullet g)$, then adding $<y, z>$ or $<z, y>$ to $Im(f \bullet g)$ must be rejected.

(ii). Each time a pair $<u, v>$ of $Im(f \bullet g)$ is modified in $<y, z>$, with either $u \neq y$ or $v \neq z$, and there is at least a $x$ in $C$ such that both $<x, u>$ and $<x, v>$ belong to $Im(f \bullet g)$, with $y \neq x$ and $y \neq z$, then modification of $<u, v>$ must be rejected.

Consequently, *MatBase* adds the pseudocode algorithms from **Figure 13** to the corresponding methods from **Figure 1**, but only when $f \bullet g$ is not symmetric and intransitive as well (case in which, according to the algorithms for the coherence and minimality of the constraint sets [2,4,7,8], inEuclideanity is redundant) and when $f \bullet g$ is not Euclidean as well.

Indeed, just like for transitivity and intransitivity, Euclideanity and inEuclideanity are orthogonal to each other, i.e., there may be relations that are both Euclidean and inEuclidean iff, for any pair $<u, v> \in$ $Im(f \bullet g)$, $u \neq v$, there is no pair $<u, w> \in Im(f \bullet g)$, $w \neq u$ (which means that any such hbfp is functional from $Im(f)$ to $Im(g)$).

According to this definition, enforcing this pair of constraints for $f \bullet g$ requires that:

(i). Each time a pair $<u, w>$ is about to be added to $Im(f \bullet g)$ and there is a pair $<u, v>$ stored by $Im(f \bullet g)$, with $u \neq v$ and $w \neq v$, then adding $<u, w>$ to $Im(f \bullet g)$ must be rejected.

(ii). Each time a pair $<x, y>$ of $Im(f \bullet g)$ is modified in $<u, w>$, with $u \neq w$, and there is at least a $v$ in $C$ such that $<u, v>$ belongs to $Im(f \bullet g)$, with $u \neq v$, then modification of $<x, y>$ must be rejected.

```
// code added to method BeforeInsert of class D
// f • g inEuclidean and not Euclidean
if not Cancel ∧ ∃z∈C such that <z, f> ∈ Im(f • g) ∧ <z, g> ∈ Im(f • g) then
    Cancel = True; display "Request rejected: f • g inEuclidean!";
end if;
// code added to method BeforeUpdate of class D
// f • g inEuclidean and not Euclidean
if not Cancel ∧ ∃z∈C such that <z, f> ∈ Im(f • g) ∧ <z, g> ∈ Im(f • g) then
    Cancel = True; display "Request rejected: f • g inEuclidean!";
end if;
```

**Figure 13.** Code added to the methods from Figure 1 when $f \bullet g$ is inEuclidean and not Euclidean.

Consequently, *MatBase* adds the pseudocode algorithms from **Figure 14** to the corresponding methods from **Figure 1** whenever $f \bullet g$ is both Euclidean and inEuclidean.

```
// code added to method BeforeInsert of class D
// f • g Euclidean and inEuclidean
if not Cancel ∧ f ≠ g ∧ ∃u∈C such that <u, g> ∈ Im(f • g), u ≠ g then
    Cancel = True; display "Request rejected: f • g both Euclidean and inEuclidean!";
end if;
// code added to method BeforeUpdate of class D
// f • g Euclidean and inEuclidean
if not Cancel ∧ f ≠ g ∧ ∃v∈C such that <f, v> ∈ Im(f • g), f ≠ v then
    Cancel = True; display "Request rejected: f • g both Euclidean and inEuclidean!";
end if;
```

**Figure 14.** Code added to the methods from Figure 1 when $f \bullet g$ is both Euclidean and inEuclidean.

In Mancas [4,28] it is also shown that any inEuclidean $f \bullet g$ which is symmetric too is intransitive and may not be connected.

## 2.11 Enforcing equivalence and null-equivalence constraints

According to the alternative definition of relation equivalence, enforcing it for $f \bullet g$ requires that $f \bullet g$ be both reflexive and Euclidean. Consequently, equivalence and null-equivalence are enforced by the algorithms from subsections 2.2 (**Figure 4**) and 2.9 (**Figure 12**).

In Mancas [4,28] it is also shown that any equivalence $f \bullet g$ which is also connected has only one equivalence class.

## 2.12 Enforcing acyclicity constraints

Generally, a hbfp $f \bullet g : D \to C^2$ is *acyclic* whenever, for any pairs $<u_1, u_2>, \ldots, <u_{n-1}, u_n> \in Im(f \bullet g)$, $n > 0$, there is no pair $<u_n, u_1> \in Im(f \bullet g)$.

According to this definition, enforcing such constraints for $f \bullet g$ requires that:

(i). Each time a pair $<x, y>$ is about to be added to $Im(f \bullet g)$ and there is a path of pairs $<y, x_1>, \ldots, <x_n, x>$, $n > 0$, exists in $Im(f \bullet g)$, then adding $<x, y>$ to $Im(f \bullet g)$ must be rejected.

(ii). Each time a pair $<u, v>$ of $Im(f \bullet g)$ is modified in $<x, y>$, with either $u \neq x$ or $v \neq y$, this must be rejected whenever a path of pairs $<y, x_1>, \ldots, <x_n, x>$, $n > 0$, exists in $Im(f \bullet g)$.

Consequently, *MatBase* adds the pseudocode algorithms from **Figure 15** to the corresponding methods from **Figure 1**, but only when $f \bullet g$ is not asymmetric and intransitive as well (case in which, according to the algorithms for the coherence and minimality of the constraint sets [2,4,7,8], acyclicity is redundant). For the pseudocode of method *IsCycle* see **Figure 2**.

In Mancas [4,28] it is also shown that any acyclic $f \bullet g$ is also asymmetric (hence, irreflexive as well); moreover, no Euclidean $f \bullet g$ may be acyclic.

```
// code added to method BeforeInsert of class D
// f • g acyclic
if not Cancel then Cancel = IsCycle("D", "x", x, "f", f, "g", g);
    if Cancel then display "Request rejected: f • g acyclic!";
end if;
// code added to method BeforeUpdate of class D
// f • g acyclic
if not Cancel then Cancel = IsCycle("D", "x", x, "f", f, "g", g);
    if Cancel then display "Request rejected: f • g acyclic!";
end if;
```

**Figure 15.** Code added to the methods from Figure 1 when $f \bullet g$ is acyclic.

## 3. Results

**Figure 16** shows the *MatBase* Algorithm *A9DH-*

*BFP* for enforcing dyadic-type hbfp constraints.

---

**MatBase Algorithm A9DHBFP for enforcing dyadic-type homogeneous binary function product constraints**

*Input*: A db software application *SA* over a set *D*, a homogeneous binary function product
  $f \bullet g : D \to (C \cup \text{NULLS})^2$, and a constraint *c* of subtype *s* on $f \bullet g$
*Output*: *SA* augmented such as to enforce *c* as well

*Strategy*: add code to the methods from Figure 1 as follows:
switch (*s*)
  case *s*: connectivity
    if $f \bullet g$ is *not* (irreflexive *and* asymmetric *or* intransitive) *then* the code from Figure 3;
    break;
  case *s*: reflexivity
    if $f \bullet g$ is *not* (null-identical *or* irreflexive *or* asymmetric *or* intransitive *or* inEuclidean *or*
      acyclic) *then* the code from Figure 4;
    break;
  case *s*: null-identity
    if $f \bullet g$ is *not* (irreflexive *or* asymmetric *or* intransitive *or* inEuclidean *or* acyclic)
      *then* the code from Figure 5;
    break;
  case *s*: irreflexivity
    if $f \bullet g$ is *not* (reflexive *or* asymmetric *or* intransitive *or* Euclidean *or* inEuclidean *or*
      acyclic) *then* the code from Figure 6;
    break;
  case *s*: symmetry
    if $f \bullet g$ is *not* (asymmetric *or* Euclidean *or* acyclic) *then* the code from Figure 7;
    break;
  case *s*: asymmetry
    if $f \bullet g$ is *not* (symmetric *or* acyclic *or* (transitive *or* Euclidean) *and* (irreflexive *or*
      intransitive)) *then* the code from Figure 8;
    break;
  case *s*: transitivity
    if $f \bullet g$ is *not* (Euclidean *or* connected *and* symmetric)
      *then* if $f \bullet g$ is intransitive *then* the code from Figure 11 *else* the code from Figure 9;
    break;
  case *s*: intransitivity
    if $f \bullet g$ is *not* (Euclidean *or* inEuclidean *and* symmetric)
      *then* if $f \bullet g$ is transitive *then* the code from Figure 11 *else* the code from Figure 10;
    break;
  case *s*: Euclideanity
    if $f \bullet g$ is *not* (acyclic *or* connected *and* symmetric)
      *then* if $f \bullet g$ is inEuclidean *then* the code from Figure 14 *else* the code from Figure 12;
    break;
  case *s*: inEuclideanity
    if $f \bullet g$ is *not* (symmetric *and* intransitive) *then*
      if $f \bullet g$ is Euclidean *then* the code from Figure 14 *else* the code from Figure 13;
    break;
  case *s*: acyclicity
    if $f \bullet g$ is *not* (Euclidean *or* reflexive *or* null-identity *or* symmetric *or* asymmetric *and*
      transitive) *then* the code from Figure 15;
    break;
  case *s*: equivalence
    if $f \bullet g$ is *not* (irreflexive *or* asymmetric *or* intransitive *or* inEuclidean *or* acyclic) *then*
      if $f \bullet g$ is *not* reflexive *then* the code from Figure 4;
      if $f \bullet g$ is *not* Euclidean *then* the code from Figure 11;
    end if;
    break;
end switch;
End MatBase Algorithm A9DHBFP;

---

**Figure 16.** *MatBase* algorithm *A9DHBFP* for enforcing dyadic-type hbfp constraints.

## 4. Discussion

For example, it is straightforward to check that applying the Algorithm *A9DHBFP* from **Figure 16** to *C = D = PERSONS* and its hbfp *Mother • Father* from section 1, *MatBase* automatically generates for *D's* class the pseudocode shown in **Figure 17**.

Generally, the Algorithm *A9DHBFP* from **Figure 16** automatically generates code that guarantees data plausibility for any hbfp for which all its properties are declared to *MatBase* as corresponding constraints, while also automatically generating appropriate data values for connectivities, reflexivities, symmetries, transitivities, and Euclideanities, thus saving most of the developing, testing, and data entering effort.

---

// class *PERSONS*

int *fOldValue, gOldValue*;
Boole *INS*;
**Method *Current*(*x, Mother, Father*)**
*fOldValue = Mother*;
*gOldValue = Father*;
*INS = False*;

**Method *BeforeInsert*(*x, Mother, Father*)**
Boole *Cancel = False*;
// *Mother • Father* inEuclidean and not Euclidean
if not *Cancel* ∧ ∃*z*∈*PERSONS* such that <*z, Mother*> ∈ *Im*(*Mother • Father*) ∧
  <*z, Father*> ∈ *Im*(*Mother • Father*) then
  *Cancel = True*; display "Request rejected: *Mother • Father* inEuclidean!";
end if;
// *Mother • Father* acyclic
if not *Cancel* then
  *Cancel = IsCycle*("*PERSONS*", "*x*", *x*, "*Mother*", *Mother*, "*Father*", *Father*);
  if *Cancel* then display "Request rejected: *Mother • Father* acyclic!";
end if;
if *Cancel* then deny inserting <*Mother, Father*> into *PERSONS*;

**Method *BeforeUpdate*(*x, Mother, Father*)**
Boole *Cancel = False*;
// *Mother • Father* inEuclidean and not Euclidean
if not *Cancel* ∧ ∃*z*∈*PERSONS* such that <*z, Mother*> ∈ *Im*(*Mother • Father*) ∧
  <*z, Father*> ∈ *Im*(*Mother • Father*) then
  *Cancel = True*; display "Request rejected: *Mother • Father* inEuclidean!";
end if;
// *Mother • Father* acyclic
if not *Cancel* then
  *Cancel = IsCycle*("*PERSONS*", "*x*", *x*, "*Mother*", *Mother*, "*Father*", *Father*);
  if *Cancel* then display "Request rejected: *Mother • Father* acyclic!";
end if;

if *Cancel* then deny saving <*Mother, Father*> to *PERSONS*;

---

**Figure 17.** *MatBase* automatically generated code in class *PERSONS* for enforcing the dyadic-type constraints on the homogeneous binary function product *Mother • Father*.

## 5. Conclusions

Not enforcing any existing business rule from the sub-universe managed by a db software application allows saving unplausible data in its db.

This paper presents the algorithms needed to enforce the dyadic-type homogeneous binary function product constraint types from the (E)MDM, which are implemented in *MatBase*, an intelligent DBMS prototype.

Moreover, as it automatically generates the corresponding code, *MatBase* is a tool of the 5th generation programming languages—*modelling as programming*: db and software architects only need to assert the properties of the hbfps (and not only, but all other (E)MDM constraint types as well), while *MatBase* saves the corresponding developing, testing, and debugging time. Obviously, these algorithms may also be used by developers not having access to *MatBase*.

## Conflict of Interest

There is no conflict of interest.

## Funding

This research received no external funding.

## Acknowledgments

This research was not sponsored by anybody and nobody other than its author contributed to it.